\documentclass[preprint,aps,superscriptaddress]{revtex4}
\pdfoutput=1
\usepackage{amsmath}
\usepackage{amsfonts}
\usepackage{grffile}
\usepackage{graphicx}
\usepackage{caption}
\usepackage{subcaption}
\usepackage{dcolumn}
\usepackage{color}
\usepackage{hyperref}
\usepackage[all]{hypcap}
\usepackage{bm}
\usepackage{color}
\usepackage{float}
\usepackage{xspace}
\usepackage{libertine}
\usepackage[makeroom]{cancel}

\newcommand{\bra}[1]{\langle #1|}
\newcommand{\ket}[1]{|#1\rangle}

\newcommand{\ketbra}[2]{| #1 \rangle \langle #2 |}
\newcommand{\Tr}[1]{\operatorname{Tr}\left[ #1 \right]}
\newcommand{\PartialTr}[2]{\operatorname{Tr}_{#1}\left[ #2 \right]}
\newcommand{\Sep}{\operatorname{Sep}}
\newcommand{\FancyH}{\mathcal{H}}
\newcommand{\Id}{\mathbb{I}\xspace}
\newcommand{\etal}{\emph{et al.}\xspace}

\begin{document}

\title{Probing Genuine Multipartite Entanglement in Large Systems}
\author{Lucas B. Vieira}
\email{lucasvb@fisica.ufmg.br}
\affiliation{Departamento de F\'{i}sica - ICEx - Universidade Federal de Minas Gerais, \\ Av. Pres. Ant\^onio Carlos 6627 - Belo Horizonte, MG, Brazil - 31270-901.}
\author{Diego L. Braga Ferreira}
\affiliation{Departamento de F\'{i}sica - ICEx - Universidade Federal de Minas Gerais, \\ Av. Pres. Ant\^onio Carlos 6627 - Belo Horizonte, MG, Brazil - 31270-901.}
\author{Thiago O. Maciel}
\affiliation{Departamento de F\'{i}sica, Universidade Federal de Santa Catarina, \\ Florianópolis, SC, Brazil - 88040-900.}
\author{Reinaldo O. Vianna}
\affiliation{Departamento de F\'{i}sica - ICEx - Universidade Federal de Minas Gerais, \\ Av. Pres. Ant\^onio Carlos 6627 - Belo Horizonte, MG, Brazil - 31270-901.}

\date{\today}

\begin{abstract}

We introduce a new kind of Entanglement Witness which is appropriate for studying genuine multipartite entanglement in large systems. The witness operator has a form that fits naturally to quantum states represented by tensor networks. It opens up novel routes for investigating properties of condensed matter systems, and gives a new perspective to understand many-body quantum correlations. We illustrate the potential of this new operator with a GHZ state, and apply the method to a transverse Ising model with the well-known  approach of Matrix Product States and Matrix Product Operators for a hundred spins. 

\end{abstract}


\maketitle

The theoretical investigation of quantum many-body systems advanced paramountly with the introduction of the Density Matrix Renormalization Group  (DMRG) \cite{whiteDensityMatrixFormulation1992}. This approach was later expressed as a variational method with the Matrix Product State (MPS) \emph{ansatz} \cite{ostlundThermodynamicLimitDensity1995,schollwockDensitymatrixRenormalizationGroup2011}. Quantum Information Theory contributed to further develop
such numerical methods \cite{verstraeteDMRGPeriodicBoundary2004},  culminating with the introduction of the Multiscale Entanglement Renormalization Ansatz (MERA) \cite{vidalClassQuantumManyBody2008}, and establishing
Tensor Network (TN) methods as the most efficient numerical approach to study quantum lattice problems (for an interesting review, see  \cite{orusTensorNetworksComplex2019}).

It is natural to expect that entanglement could shed light on  the analysis of quantum complex systems. In a synergy of Quantum Information Theory and Condensed Matter Physics, 
such expectation was put into practice. Bipartite entanglement has been used successfully in the characterization of quantum phase transitions, as can be seen in seminal works
by Osborne and Nielsen  \cite{osborneEntanglementSimpleQuantum2002}, or Vidal \etal \cite{vidalEntanglementQuantumCritical2003}, where one-dimensional chains of spin-1/2 particles were studied. Since then, a myriad of papers have been published using such approach.

For the present discussion, the relevant point is that all such works so far  dealt  exclusively with bipartite entanglement, which, for pure states, is easy to compute. In this case, one just needs the reduced density matrix corresponding to a subsystem, say one or a few particles, which is efficiently computed by TN approaches. Access to higher order correlations like multipartite entanglement might unfold properties oblivious to bipartite entanglement. To detect these properties one must first know how to compute multipartite entanglement, which is not possible by means of local reduced density matrices, as in the bipartite case. Ideal assessment of multipartite entanglement demands witness operators \cite{brandaoSeparableMultipartiteMixed2004, brandaoRobustSemidefiniteProgramming2004, brandaoWitnessedEntanglement2006, ieminiQuantifyingQuantumCorrelations2013}, which do not correspond to local properties. In this letter, we introduce a multipartite entanglement witness (EW) operator capable of performing this task and which is well suited for tensor networks.

\begin{figure}[H]
	\centering
	\includegraphics[width=0.6\linewidth]{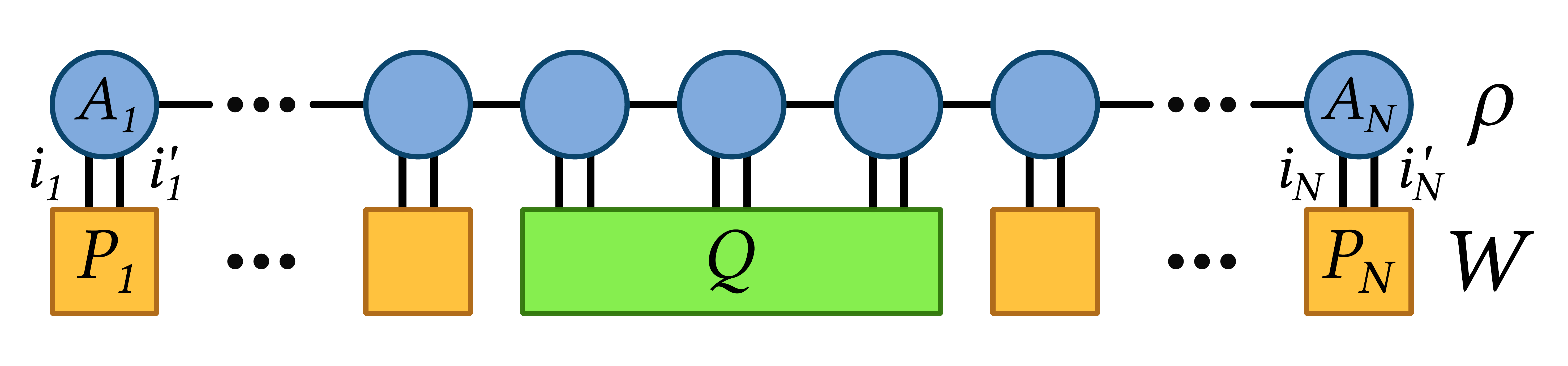}
	\caption{A tensor network employing our proposed witness operator. The state $\rho$ and the witness $W$ are both represented as Matrix Product Operators~\cite{orusTensorNetworksComplex2019}. $W$ contains a smaller witness operator $Q$ acting on multiple sites, which can be placed anywhere along the chain, padded with left and right positive semidefinite operators $P$. This witness $Q$ acts as a window within which multipartite entanglement may be probed along the chain.}
	\label{fig:wnetwork}
\end{figure}

Consider a very large chain of particles and a ``window'' that can slide over the chain. We want to evaluate the multipartite entanglement of the particles within this window. Note that this is different from taking a partial trace, ``cutting'' the window from the chain, and analyzing the corresponding reduced density matrix (\hyperref[fig:wnetwork]{Fig.~\ref*{fig:wnetwork}}). 
The influence of particles outside the window is accounted for by positive semidefinite operators $P \neq \Id$, such that the operator for the whole chain is defined as
\begin{equation}\label{eq:rov1}
W=P_L \otimes Q \otimes P_R,
\end{equation}
where $P_L$ and $P_R$ are positive semidefinite operators on the Hilbert spaces corresponding, respectively, to the left ($\FancyH_L$) and right  ($\FancyH_R$) of the window, and $Q$ is a genuine multipartite witness operator acting on the Hilbert space inside of it ($\FancyH_\text{w}$). For a relatively small
window, we can use an optimal entanglement witness $Q$ constructed with the successful methodologies previously developed \cite{brandaoSeparableMultipartiteMixed2004, brandaoRobustSemidefiniteProgramming2004, brandaoWitnessedEntanglement2006, ieminiQuantifyingQuantumCorrelations2013,jungnitschTamingMultiparticleEntanglement2011}.

The expectation value of an EW must be non-negative for all separable states, and negative for some non-separable (entangled) state. Let $D(\FancyH)$  and $\Sep{}(\FancyH)$ (with $\Sep{}(\FancyH)\subset D(\FancyH)$) be the sets of density operators and separable states  on the Hilbert space
$\FancyH$, respectively. As $Q$ is an EW by hypothesis, then $\Tr{Q \sigma_\text{w}} \geq 0$, $\forall \, \sigma_\text{w} \in \Sep{}(\FancyH_\text{w})$. Now it is straightforward to check that $W$ is an EW. As any separable state can be decomposed as convex combination of pure separable states,  and as  $\bra{\psi}W\ket{\psi} \geq 0$, for  $\ket{\psi}=\ket{\psi_L} \ket{\psi_\text{w}} \ket{\psi_R}$, 
with $\ket{\psi_L}\in \FancyH_L$,  $\ket{\psi_\text{w}} \in \FancyH_\text{w}$ and $\ket{\psi_R} \in \FancyH_R$, thus 
$\Tr{W\sigma} \geq 0, \forall \, \sigma \in \Sep{}(\FancyH_L \otimes \FancyH_\text{w} \otimes \FancyH_R)$, by convexity.

We now show that $W$ (\hyperref[eq:rov1]{Eq.~(\ref*{eq:rov1})}) is an interesting witness operator and that the addition of $P$ acts as a contrast, in the sense that it highlights the entanglement inside the window, if well chosen.
Taking $P$ as the identity operator ($\Id = \Id_L \otimes \Id_R$) is equivalent to ignoring the chain outside the window, by tracing it out:  $\Tr{W\rho} = \Tr{ (\Id_L\otimes Q \otimes \Id_R) \, \rho} = \Tr{Q \PartialTr{LR}{\rho}} = \Tr{Q \rho_\text{w}}$,
where $\rho_\text{w}$ is the reduced density matrix corresponding to the window. The only entanglement that can be inferred by means of $\rho_\text{w}$ is the bipartite one between the window and 
the rest of the chain. The residual entanglement in $\rho_\text{w}$, if any, is not the one the particles in $\FancyH_\text{w}$ had when they were part of the whole system ($\FancyH_L \otimes \FancyH_\text{w} \otimes \FancyH_R$). To make this point clear, consider the whole chain formed by $N$ qubits, in the maximally entangled GHZ state:
\begin{equation}\label{eq:rov2}
\ket{\text{GHZ}_N} = \frac{\ket{0}^{\otimes N} + \ket{1}^{\otimes N}}{\sqrt{2}}, \qquad \rho_{\text{GHZ}_N} = \ket{\text{GHZ}_N}\bra{\text{GHZ}_N}.
\end{equation}
Any reduced density matrix of the GHZ state is separable, therefore the EW with the contrast operator $P$ set to the identity informs us nothing about the entanglement in the chain. On the other hand, consider a window of $M$ qubits length ($M<N$). Taking as the contrast operator $P = \ket{\text{GHZ}_{N-M}}\bra{\text{GHZ}_{N-M}}$, we obtain
\begin{equation}\label{eq:rov3}
\Tr{W \rho_{\text{GHZ}_N}}=\frac{1}{2} \Tr{Q\rho_{\text{GHZ}_M}},
\end{equation}
and thus we have witnessed that the particles inside the window are indeed maximally entangled. Notice that the window can have any size and be placed anywhere, therefore we have characterized the true multipartite entanglement for the whole chain, according to Q.

For the sake of computational efficiency in TN applications, a contrast operator formed by one-particle operators would be ideal.  Taking the state $\ket{P_1} = \tfrac{1}{\sqrt{2}}(\ket{0} + \ket{1})$,  and forming the positive operator $P_1=2\ketbra{P_1}{P_1}$,  we
obtain the same result as before, except for a multiplicative factor, namely:
\begin{equation}\label{eq:rov4}
\Tr{ (Q \otimes P_1^{\otimes N-M}) \, \rho_{\text{GHZ}_N}} = \Tr{Q \rho_{\text{GHZ}_M}}.
\end{equation} 

The very simple structure of the GHZ state allowed us to derive the optimal contrast operator ($P$) straightforwardly. For an arbitrary state, however, $P$ should be optimized numerically. In the case of translational symmetry (a desirable property in many TN applications) one can see that the contrast operators may be all made equal, as in the GHZ example. In the absence of symmetry, optimization becomes non-trivial due to non-convexity of the problem.

\medskip

Assuming that $P$ is a projector, we can interpret this operator as part of the following map:
\begin{equation}\label{eq:rov5}
\alpha \tilde{\rho} = (P_L\otimes \Id_\text{w} \otimes P_R) \, \rho \, (P_L \otimes \Id_\text{w} \otimes P_R), \quad \text{with} \; \Tr{\tilde{\rho}} = 1, \text{and} \; 0 \le \alpha \le 1. 
\end{equation}
This implies $\Tr{W \rho} = \alpha \Tr{W \tilde{\rho}} = \alpha \Tr{Q \tilde{\rho}_\text{w}}$, where $\tilde{\rho}_\text{w}$ is the reduced density matrix of $\tilde{\rho}$ corresponding to the window. Therefore, the optimal contrast operator minimizes $\Tr{W \rho}$ by minimizing $\Tr{Q \tilde{\rho}_\text{w}}$. In practice, the sub-normalization after projections must be accounted for to ensure numerical stability over large chains.

\medskip

\begin{figure}[H]
     \centering
     \begin{subfigure}[b]{0.475\textwidth}
         \centering
         \includegraphics[width=\textwidth]{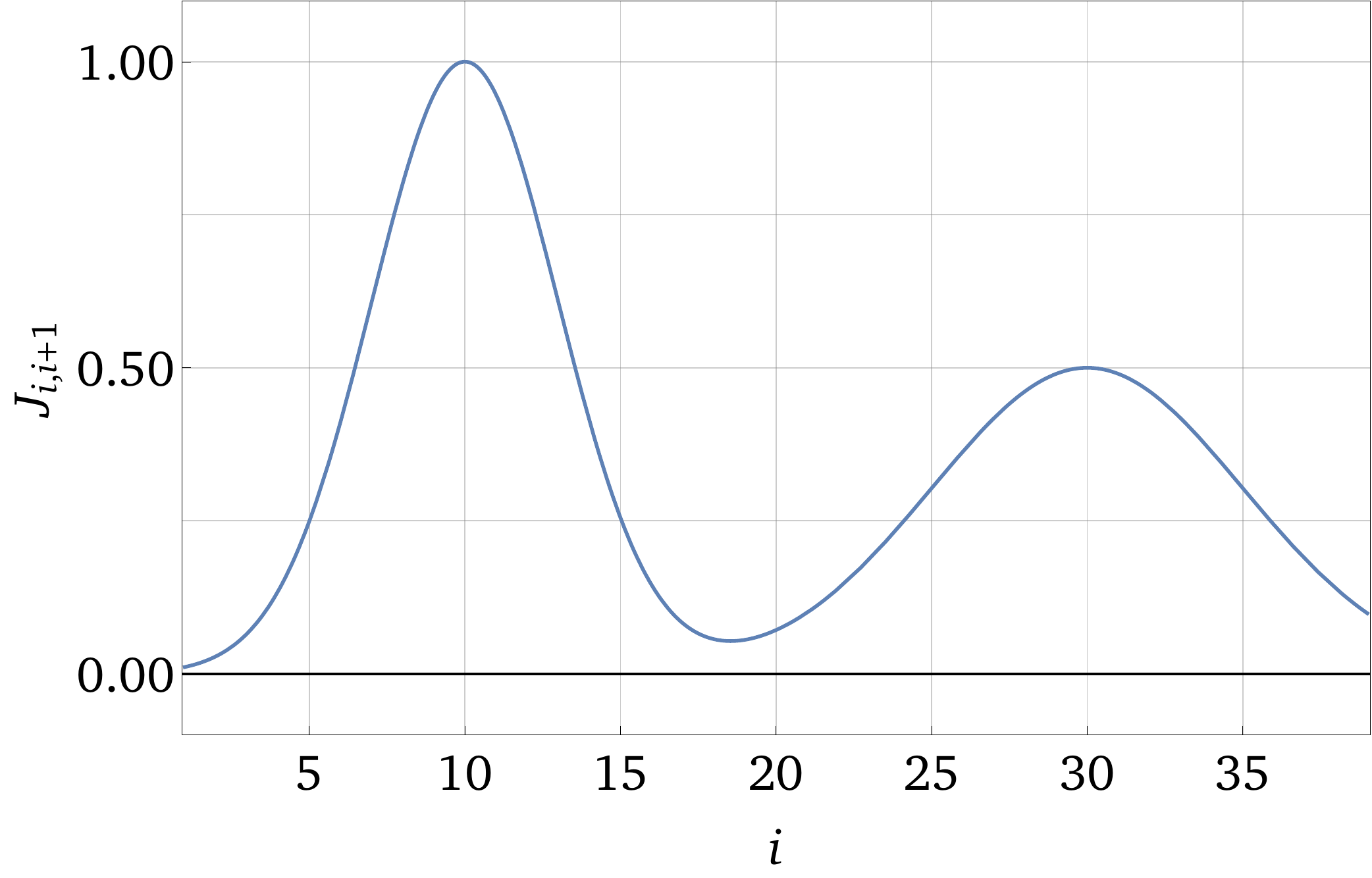}
         \caption{Interaction strength per site.}
         \label{fig:doubleisingspikes}
     \end{subfigure}
     \hfill
     \begin{subfigure}[b]{0.475\textwidth}
         \centering
         \includegraphics[width=\textwidth]{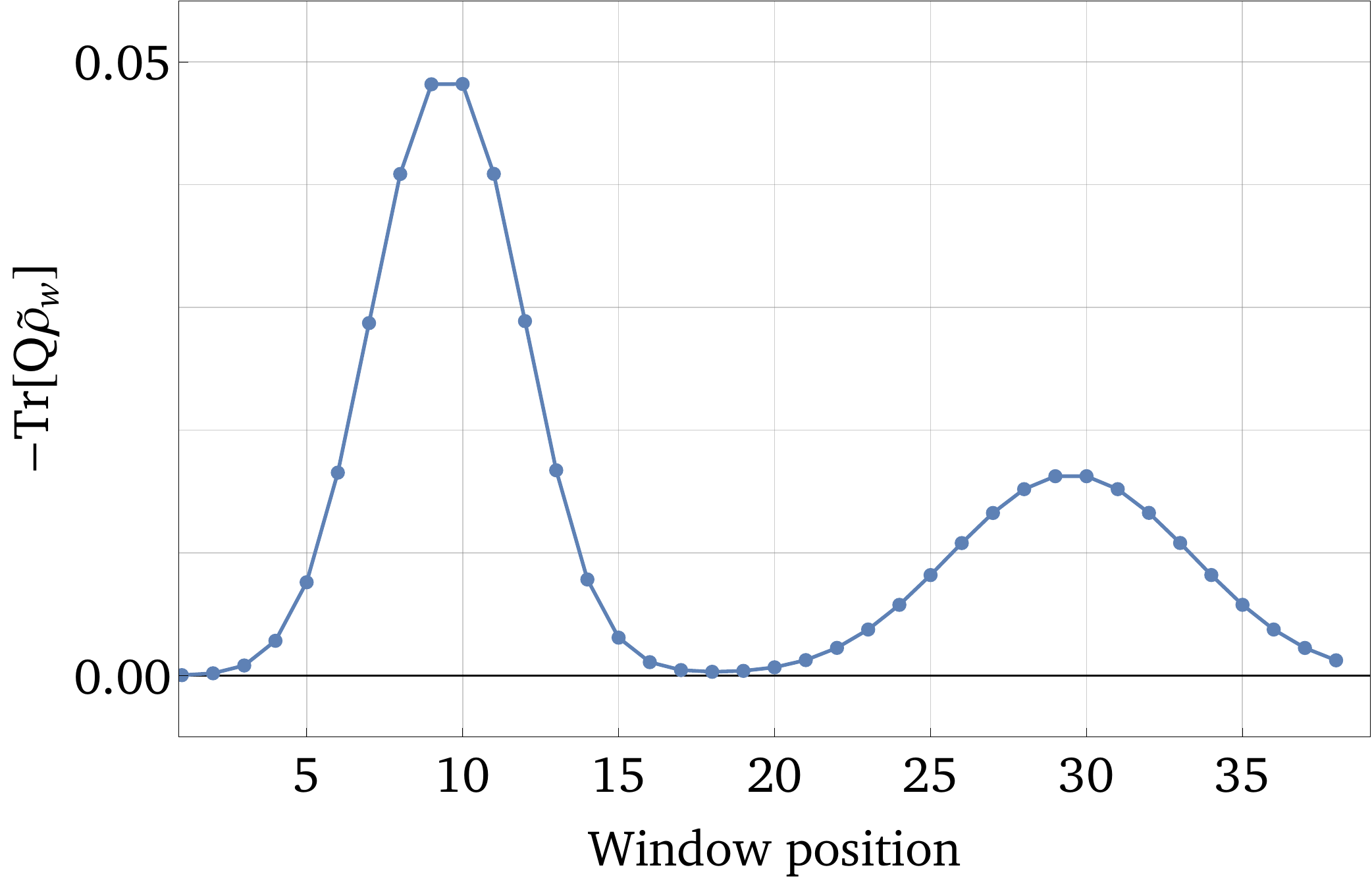}
         \caption{Genuine 3-partite entanglement.}
         \label{fig:doubleisingspikeswitness}
     \end{subfigure}
     \caption{Entanglement dependency on the window position. While our witness senses the entire Hilbert space, it still allows for local entanglement probing. (Position is taken to be the leftmost site position within the window, hence the small offset.)}
\end{figure}

We will now demonstrate the usefulness of our operator in more practical scenarios. Let us consider a finite chain of $N$ spin-1/2 particles interacting under the transverse Ising Hamiltonian,
\begin{equation}\label{eq:isingh}
	\FancyH = - \sum_{i=1}^{N-1} J_{i,i+1} \sigma^\text{z}_i \sigma^\text{z}_{i+1} - g \sum_{i=1}^N \sigma^\text{x}_i,
\end{equation}
where $g$ represents the intensity of the transverse field and $J_{i,i+1}$ a nearest-neighbor interaction strength. To illustrate the entanglement probing features of our witness consider a chain of $N = 40$ spins and a non-uniform interaction term,
\begin{equation}\label{eq:Jdist}
	J_{i,i+1} = \text{e}^{-\frac{(i - x_a)^2}{2 a^2}} + \tfrac{1}{2} \text{e}^{-\frac{(i - x_b)^2}{2 b^2}},
\end{equation}
with $x_a = 10$, $a = 3$, $x_b = 30$, and $b = 5$. This profile is illustrated in \hyperref[fig:doubleisingspikes]{Fig.~\ref*{fig:doubleisingspikes}}. A ground state of this Hamiltonian was obtained using DMRG~\cite{schollwockDensitymatrixRenormalizationGroup2011,ITensorIntelligentTensor}, with $g = 1.1$, which was subsequently probed at each location in the chain using a window width of 3 sites. Numerical results are shown in \hyperref[fig:doubleisingspikeswitness]{Fig.~\ref*{fig:doubleisingspikeswitness}}, where it is clear that our proposed operator effectively acts as a genuine multipartite \emph{entanglement probe} for sections of the entire chain.

For an even more illustrative example, let us take the same Hamiltonian from \hyperref[eq:isingh]{Eq.~(\ref*{eq:isingh})} now with $J_{i,i+1} = 1$. This is the familiar transverse Ising model, which has a known phase transition at $g = 1$ in the thermodynamic limit. We will probe the genuine 3-partite entanglement at the center of a chain for varying $g$, for many chain lengths, and compare our method with a witness applied to a reduced density matrix.
\begin{figure}[H]
     \centering
     \begin{subfigure}[b]{0.475\textwidth}
         \centering
         \includegraphics[width=\textwidth]{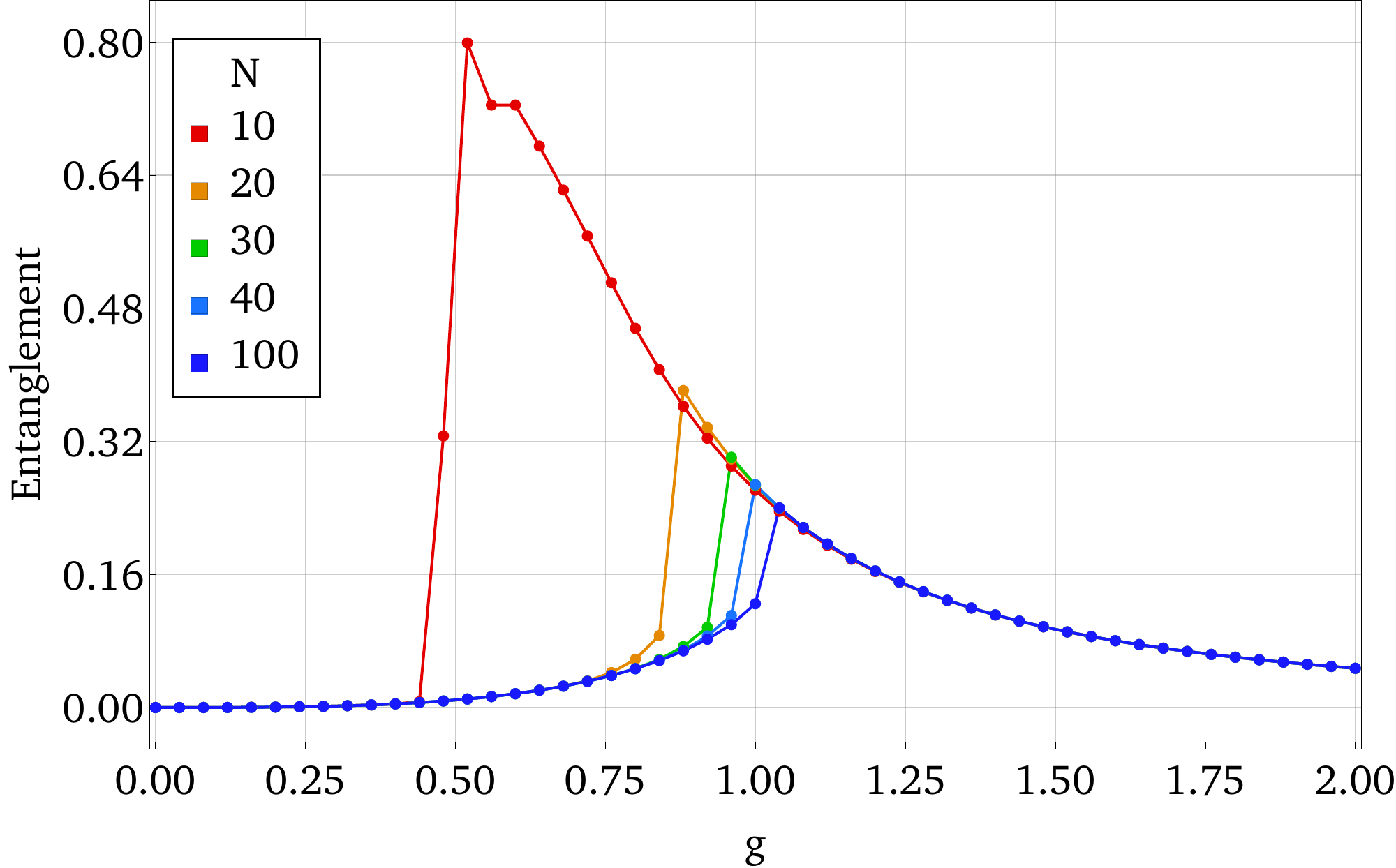}
         \caption{Witness on the entire chain ($\tilde{\rho}_\text{w}$).}
         \label{fig:isingtransition}
     \end{subfigure}
     \hfill
     \begin{subfigure}[b]{0.475\textwidth}
         \centering
         \includegraphics[width=\textwidth]{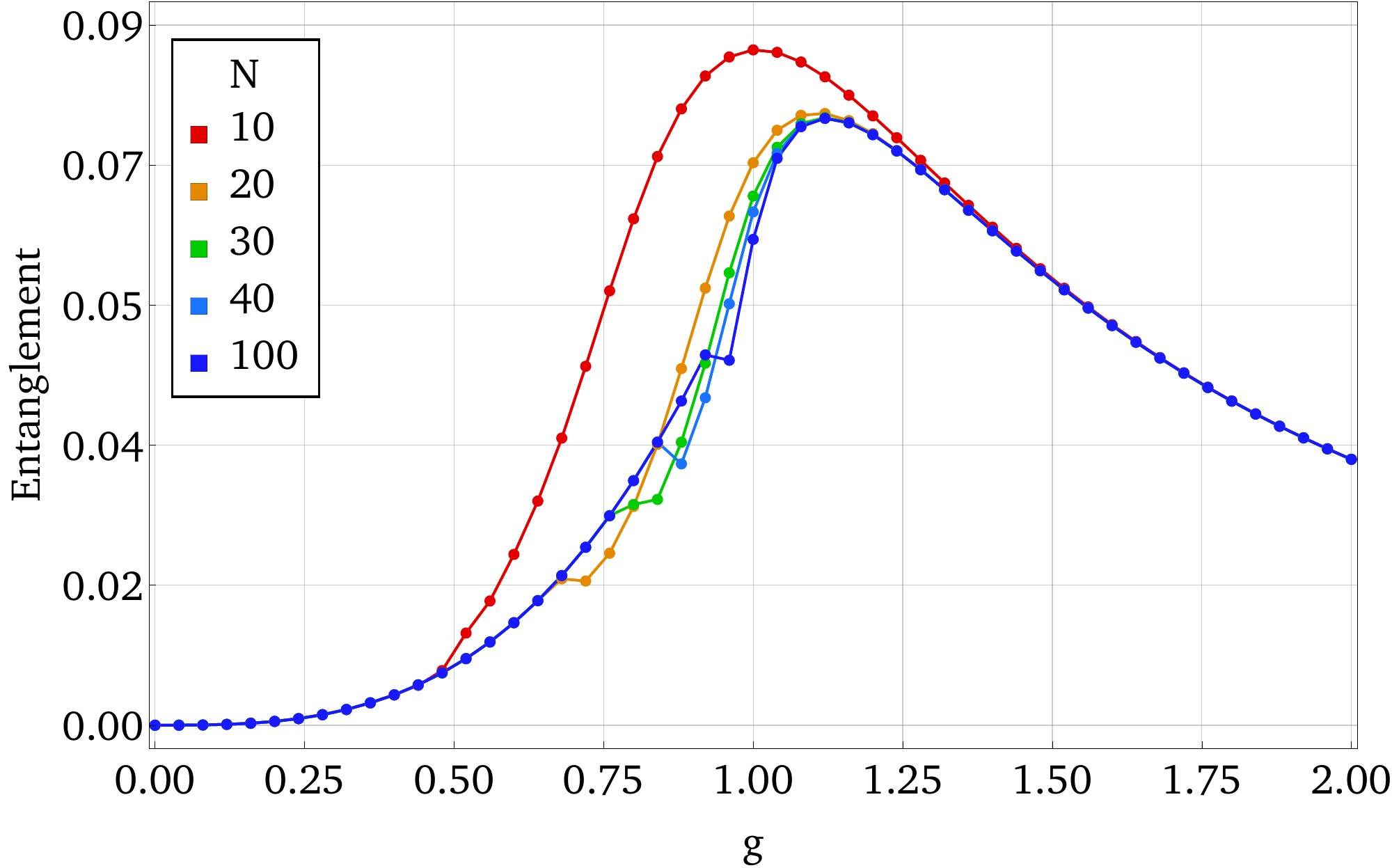}
         \caption{Witness on the reduced density matrix ($\rho_\text{w}$).}
         \label{fig:isingtransitiontrace}
     \end{subfigure}
	\caption{Genuine 3-partite witnessed entanglement~\cite{brandaoWitnessedEntanglement2006} at the center of a transverse Ising model chain, as measured by our witness (using $\Tr{Q \tilde{\rho}_\text{w}}$) \emph{vs.} using the reduced density matrix on three sites. The location of the transition shifting towards $g = 1$ as $N$ increases is a known and expected result, 
	which our witness seems to display. Using the reduced density matrix the dependency on system size is much less evident.}
	\label{fig:isingtransition}
\end{figure}

We have demonstrated how it is possible to witness the multipartite entanglement in a large system. The results obtained so far required good initial guesses for the contrast operators $P$ before optimization. The ideal method for finding the optimal family of operators for a given state remains an open problem under investigation.

\acknowledgments We are grateful to Marcello Nery, Carlos H. Monken, and Douglas S. Gon\c{c}alves for fruitful discussions. We acknowledge financial support from the Brazilian agencies FAPEMIG, CAPES, and CNPq INCT-IQ (465469/2014-0).

\bibliography{bibliography}
\bibliographystyle{ieeetr}

\end{document}